\documentclass[11pt]{article}
\pdfoutput=1

\usepackage{geometry}

\usepackage{amsmath,amsfonts}
\usepackage{graphicx}
\usepackage{tensor}
\usepackage{subfig}
\usepackage{hyperref}

\def\subfigure{\subfloat}
\numberwithin{equation}{section}

\newcommand{\nn}{\nonumber}

\def\eqa{\begin{eqnarray}}
\def\eqae{\end{eqnarray}}
\def\eq{\begin{equation}}
\def\eqe{\end{equation}}
\def\be{\begin{equation}}
\def\ee{\end{equation}}
\def\bea{\begin{eqnarray}}
\def\eea{\end{eqnarray}}
\def\ba{\begin{array}}
\def\ea{\end{array}}
\def\bd{\begin{displaymath}}
\def\ed{\end{displaymath}}

\def\Tr{{\rm Tr}}
\def\tr{{\rm tr}}
\def\>{\rangle}
\def\<{\langle}

\begin{document}

\begin{titlepage}
\hfill MCTP-17-12
\vspace{1cm}
\begin{center}

{\Large \textbf{Comments on Higher Rank Wilson Loops in ${\cal N}=2^*$}}\\[4em]

\renewcommand{\thefootnote}{\fnsymbol{footnote}}

{\large  James T. Liu${}^a$,  Leopoldo A.~Pando Zayas${}^{b}$ and Shan Zhou${}^{c}$}\\[3em]

\renewcommand{\thefootnote}{\arabic{footnote}}

${}^{a,b}$\emph{Michigan Center for Theoretical Physics,  Randall Laboratory of Physics\\ The University of
Michigan,  Ann Arbor, MI 48109, USA}\\[1em]

${}^b$\emph{The Abdus Salam International Centre for Theoretical Physics\\ Strada Costiera 11,  34014 Trieste, Italy}\\[1em] 

${}^c$\emph{Institute for Interdisciplinary Information Sciences \\ Tsinghua University, Beijing 100084, China}\\[6em]

\abstract{For  ${\cal N}=2^*$ theory with $U(N)$ gauge group we evaluate expectation values of Wilson loops in representations described by a rectangular Young tableau with $n$ rows and $k$ columns. The evaluation reduces to a two-matrix model and we explain, using a combination of numerical and analytical techniques, the general properties of the eigenvalue distributions in various regimes of parameters $(N,\lambda,n,k)$ where $\lambda$ is the 't Hooft coupling. In the large $N$ limit we present analytic results for the leading and sub-leading contributions. In the particular cases of only one row or one column we reproduce previously known results for the totally symmetry and totally antisymmetric representations. We also extensively discusss the  ${\cal N}=4$ limit of  the ${\cal N}=2^*$ theory. While establishing these connections we clarify aspects of various orders of limits and how to relax them; we also find it useful to explicitly address details of the genus expansion. As a result, for the totally symmetric Wilson loop we find new contributions that improve the comparison with the dual holographic computation at one loop order in the appropriate regime. }

\end{center}

\end{titlepage}


\section{Introduction}

Wilson loops are non-local operators in gauge theory; they serve as order parameters in many situations. In the context of the AdS/CFT correspondence Wilson loops play a particularly important role as they are described, at leading order, by classical configurations of strings and branes \cite{Maldacena:1998im,Rey:1998ik}. These classical configurations represent a controlled departure from the strict supergravity limit into stringy aspects of the correspondence. Indeed, the AdS/CFT dictionary has been enlarged to include D3 and D5 branes  corresponding to Wilson loops in the symmetric and antisymmetric representations of $SU(N)$ for ${\cal N}=4$ SYM  \cite{Drukker:2005kx,Gomis:2006sb,Gomis:2006im,Yamaguchi:2006tq,Hartnoll:2006is}.

More recently, due to the advent of localization techniques \cite{Pestun:2007rz}, the expectation values of some supersymmetric Wilson loops have been expressed as matrix models. This situation sets the stage, in the context of the AdS/CFT correspondence, for a very rich interplay (for a recent review see \cite{Zarembo:2016bbk}).

In this manuscript we study the expectation values of certain Wilson loops in a supersymmetric deformation of  ${\cal N}=4$ supersymmetric Yang Mills known as ${\cal N}=2^*$ theory. One of our main driving motivations is to consider the prototypical dual pair in AdS/CFT, namely, the equivalence of string theory on $AdS_5\times S^5$ with $N$ units of RR five-form flux and ${\cal N}=4$ SYM in a less symmetric corner.  The hope is to extract lessons about non-conformal theory where we expect interesting aspects of gauge dynamics  play an important role, particularly when results are no longer determined by symmetries as could be arguably the case in many situations  in the prototypical $AdS_5\times S^5/{\cal N}=4$ SYM case.

There is already a number of works which precisely address various properties of the AdS/CFT correspondence for the case of ${\cal N}=2^*$ theory. For example, the intricate structure of phases in ${\cal N}=2^*$ was discussed in  \cite{Russo:2013qaa,Zarembo:2014ooa}. Initial discussions on the holographic  side were presented in  \cite{Buchel:2013id,Bobev:2013cja}. Along the lines that we pursue in this paper there has already been a number of papers evaluating expectation values of some supersymmetric Wilson loops in ${\cal N}=2^*$  \cite{Chen-Lin:2015dfa,Chen-Lin:2015xlh}. Finally, a very recent test going beyond the leading order and delving into quantum corrections on the holographic side  \cite{Chen-Lin:2017pay} paves the way for more precision holography in this context.

In this paper we consider the  ${\cal N}=2^*$ theory with $U(N)$ gauge group and evaluate expectation values of Wilson loops in representations described by rectangular Young tableaux with $n$ rows and $k$ columns. Our computation of the vacuum expectation value of the rectangular Wilson loop relies on two key approximations. First, we assume that the leading order answer is given by a saddle point evaluation which requires the large-$N$ limit. Second, we assume that the characteristic eigenvalue distribution is given by two groups of eigenvalues which are widely separated with separation $k \lambda/(4N)$. Thus, for large separation we require large $k$ or large $\lambda$. We  provide a systematic way of introducing corrections to this leading order approximation in various parameters. To carefully account for all the corrections, we are forced, in some situations, to investigate aspects of the genus expansion.

The rest of the paper is organized as follows. In section \ref{Sec:Prelim} we briefly review the ${\cal N}=2^*$  theory and describe the general computational setup. Section \ref{Sec:SP} discusses the saddle point approximation to the expectation value of the Wilson loops and highlights generic properties. In section \ref{Sec:SandA} we discuss how our computation relates to the  totally symmetric and totally antisymmetric cases that have been discussed in the literature. Section \ref{Sec:OneLoopCorrections} is devoted to the one-loop corrections; we discuss the general case and revisit the totally symmetric case in detail. We conclude in section \ref{Sec:Conclusions}.  In appendix \ref{App:genus} we provide some details of the systematic genus expansion that are used in the main text and present a pedagogically intructive example. Appendix \ref{App:Ak} revisits previously overlooked details of the computation of the Wilson loop in the totally antisymmetric representation.

\section{Wilson loops in ${\cal N}=2^*$ theory}\label{Sec:Prelim}%

The ${\cal N} = 2^*$ theory is a relevant perturbation of maximally supersymmetric ${\cal N} = 4$ SYM by a
combination of dimension two and dimension three operators that preserves half of the supersymmetry.
The field content of ${\cal N} = 2^*$ is the same as in the ${\cal N} = 4$ theory. In addition to the gauge field, there
are six scalars which are typically denoted by $\Phi_1, \ldots , \Phi_4, \Phi, \Phi'$  and four Majorana fermions. As in ${\cal N}=4$, all fields are in the adjoint representation of the gauge group. The relevant
perturbation adds equal masses to $\Phi_I$ and their superpartners; this mass scale is denoted by $M_0$. In summary, in ${\cal N}=2$ language one has a vector multiplet and a massive hypermultiplet. See  \cite{Pestun:2007rz,Bobev:2013cja} for a presentation of the ${\cal N}=2^*$ theory with details pertaining to its definition on $S^4$.

It is quite remarkable that localization techniques can be used to compute a host of supersymmetric observables for field theories \cite{Pestun:2007rz}. Particularly relevant for us are certain Wilson loops  in the ${\cal N}=2^*$ theory  defined as
\begin{equation}
	W_\mathcal{R}(C)=\left<\tr_\mathcal{R}P \exp\left[\int_C ds(iA_\mu \dot{x}^\mu+\Phi|\dot{x}|)\right]\right>,
\end{equation}
where $\mathcal{R}$ is an arbitrary representation of $U(N)$ and $\Phi$ is a scalar from the vector multiplet. The main result of \cite{Pestun:2007rz} effectively turns expectation values of supersymmetric Wilson loops, whose contour is the large circle on $S^4$,  in some four dimensional ${\cal N}=2$ supersymmetric field  theories into matrix integrals. Namely, 

\begin{align}
\langle W_{\cal R}(C)\rangle &= \frac{1}{Z_{S^4}}\frac{1}{{\rm Vol}(G)}\int\limits_{\mathfrak{g}} [dM]e^{-\frac{8\pi^2 R^2 }{g^2_{YM}}(M,M)}Z_{1-loop}(iM)|Z_{inst}(iM, R^{-1}, R^{-1},q)|^2 {\rm Tr}_{\cal R}e^{2\pi iRM}, \nonumber \\
Z_{S^4}&=\frac{1}{{\rm Vol}(G)}\int\limits_{\mathfrak{g}} [dM]e^{-\frac{8\pi^2 R^2 }{g^2_{YM}}(M,M)}Z_{1-loop}(iM)|Z_{inst}(iM, R^{-1}, R^{-1},q)|^2 .
\end{align}
where $R$ is the radius of $S^4$, and $dM$ represents integration over hermitian matrices with the Haar measure. More practically, for gauge invariant observables, the integral over the Lie algebra $\mathfrak{g}$ may be turned into an integral over its Cartan subalgebra. 
The instanton partition function is the generating function of instantons of a given topological charge; that is, it is a sum with coefficients $q^n=\exp(2\pi i  n \tau)$ where $\tau$ is the complexified Yang-Mills coupling $\tau =\theta/2\pi +i 4\pi/g_{YM}^2$. Therefore, in the large-$N$ limit with a fixed  't Hooft coupling, $\lambda$, the instanton contribution is exponentially suppressed $e^{-8\pi^2 N/\lambda}$. Thus,  by working in the large-$N$ limit, we may consistently set  $Z_{inst}=1$.

Using standard matrix model techniques \cite{Eynard:2015aea} one reduces the $\mathcal N=2^*$ problem to an integration over eigenvalues, $m_i$:
\begin{equation}
\label{Eq:PartFn}
	\mathcal{Z}=\int dm\prod_{i<j}\mathcal{Z}_{1-\mathrm{loop}}(m_i-m_j,M_0,R)e^{-\frac{8\pi^2NR^2}{\lambda}\sum m_i^2},
\end{equation}
where the one-loop contribution is
\begin{equation}
	\mathcal{Z}_{1-\mathrm{loop}}(x,M_0,R)=\frac{x^2H^2(x,R)}{H(x+M_0,R)H(x-M_0,R)}, \quad  H(u,R)=\prod_{n=1}^\infty\left(1+\frac{R^2u^2}{n^2}\right)^ne^{-\frac{R^2u^2}{n}}.
\end{equation}
Note that, in the large $N$ limit, the difference between $SU(N)$ and $U(N)$ group is suppressed. So from now on we do not distinguish between these two gauge groups.

The Wilson loop expectation value is obtained by evaluating
\begin{equation}
	W_\mathcal{R}(C)=\left<\tr_\mathcal{R}e^{LM_0}\right>,
\end{equation}
where $L=2\pi R$ is the length of the contour $C$ and 
\begin{equation}
	M=\mathrm{diag}(m_1,\dots,m_N).
\end{equation}
After the scaling \(m_i\to \left.m_i\right/L\), the partition function only depends on the product $M_0R$ through \(\mathcal{Z}_{1-\text{loop}}\), up to a proportionality constant which will be canceled when evaluating the vacuum expectation value (vev). Therefore, it is obvious that in the limit \(M_0 R\to 0\), the \(\mathcal{N}=2^*\) theory simply reduces to \(\mathcal{N}=4\) theory, and we will use this limit to compare with results present in the literature. The general rule relating the results before and after the scaling is: set \(R=1/2\pi\) and then  \(M_0\mapsto 2\pi  M_0 R\).

We will focus our analysis on higher dimensional representations following an approach first discussed, to our knowledge, by Okuda  \cite{Okuda:2007kh} for the case of ${\cal N}=4 $ SYM. Some important technical aspects of this approach were  also used in \cite{Halmagyi:2007rw}.  One significant  result within this approach is the computation of Wilson loops in an arbitrary representation which is equivalent to producing a spectral curve from a given Young tableau. On the holographic side the back-reacted geometry corresponding to a Wilson loop in arbitrary representations was constructed in \cite{DHoker:2007mci}. Indeed, in \cite{Okuda:2008px} the authors elaborated on the method of \cite{Okuda:2007kh} with the  goal of comparing with the holographic side. 

The key identities that we are going to use pertain to forms of writing the trace over a representation ${\cal R}$. They are discussed in \cite{Okuda:2007kh} in the context of computing expectation values  of Wilson loops in ${\cal N}=4$ SYM but are, of course, well known statements in group theory \cite{fulton1991representation}:
\bea
\label{2UMatrix}{\rm Tr}_{\cal R} e^M&=&\int dU  \det(1+e^M\otimes U^{-1})\,\,{\rm Tr}_{{\cal R}^T}U, \\
\label{2VMatrix}{\rm Tr}_{\cal R} e^M&=&\int dV\frac{1}{\det(1-V^{-1}\otimes e^M)}\,\,{\rm Tr}_{{\cal R}}V, 
\eea
where $U$ and $V$ are unitary matrices, $dU$ and $dV$ denote  the Haar measure and ${\cal R}^T$ stands for the transpose of ${\cal R}$.

In this manuscript we specialize to rectangular Young tableaux with $n$ rows and $k$ columns. It is, therefore, natural to consider the matrix $U$ to be  a $k\times k$  unitary matrix and the $V$ matrix to be an $n\times n$ one.  In this case we have that
\be
{\rm Tr}_{{\cal R}^T}=(\det  U)^n, \qquad {\rm Tr}_{\cal R}V=(\det V)^k.
\ee
Therefore the starting expressions for the Wilson loop observables are 
\begin{align}
\label{eq:UVmat}
\langle W_{\cal R}\rangle&= \frac{1}{\mathcal{Z}}\int dM dU \exp\left(-S(M)_{{\cal N}=2^*}\right) \det(1+e^M\otimes U^{-1}) (\det  U)^n , \nonumber \\
\langle W_{\cal R}\rangle&= \frac{1}{\mathcal{Z}}\int dM dV \exp\left(-S(M)_{{\cal N}=2^*}\right) 
\frac{1}{\det(1-V^{-1}\otimes e^M)}(\det  V)^k,
\end{align}
where $S(M)_{{\cal N}=2^*}$ can be read off from the partition function given in Eq.~(\ref{Eq:PartFn}).
Note that the $U$ matrix and $V$ matrix expressions are formally equivalent, so whichever one is more convenient can be used.  However, the results may not be identical once the saddle point and other approximations are performed.

\section{Saddle point approximation: general properties}\label{Sec:SP}

The vacuum expectation values of the Wilson loops we study here have  a vast parameter space $(N,M_0R,\lambda, k,n)$. We are going to consider always the large-$N$ limit and explore various regimes in the rest of the parameters. We are going to follow standard matrix model techniques for obtaining the expectation values of the Wilson loop. Operationally, large $N$ means that we are going to focus on the saddle point approximation. 

In this section we show some of the key properties of the eigenvalue distribution. A  number of these properties, like the separation of eigenvalues in two groups, were first established for the ${\cal N}=4$ case.  Our goal is to present various analytical results in the computation of the expectation values of rectangular Wilson loops. However, to confirm some of the results and to develop our intuition we will also conduct some numerical explorations.

\subsection{Rectangular Wilson loop}

We will focus on the rectangular representation $\mathcal{R}$ with $n$ rows and $k$ columns. Collecting various results from the previous sections, and focusing on the $U$ matrix in (\ref{eq:UVmat}),  one can write the vev of such a Wilson loop  as
\begin{equation}\label{WLwithUMatrix}
\begin{split}
	\left<W_\mathcal{R}\right> = \frac{1}{k!(2\pi)^k}\frac{1}{\cal Z}\int\prod_{a=1}^k du_a \prod_{i=1}^N dm_i \exp\left[-\frac{2N}{\lambda} \sum_{i=1}^N m_i^2 +\sum_{i<j}\log \mathcal{Z}_{1-\mathrm{loop}}(m_i-m_j)+n\sum_{a=1}^k u_a\right. \\
	 \left.+\sum_{a<b}\log\left(2\sinh\frac{u_a-u_b}{2}\right)^2+\sum_{a,i}\log(1-e^{m_i-u_a})\right],
\end{split}
\end{equation}
where 
\begin{equation}
	\mathcal{Z}_{1-\mathrm{loop}}(x)\equiv\mathcal{Z}_{1-\mathrm{loop}}(x,2\pi M_0R,1/2\pi).
\end{equation}
Since the radius $R$ can be scaled away, we set $R=1$ (i.e.\ we take $M_0R\to M_0$, and the decompactification limit $R\gg1$ is replaced with $M_0\gg1$). When $x$ is large, the one-loop function is simply a re-scaling of the Vandermonde determinant in $\mathcal{N}=4$ theory, namely:
\begin{equation}
	\log\mathcal{Z}_{1-\mathrm{loop}}(x)\to2(1+M_0^2)\log|x|,\quad |x|\gg2\pi M_0.
\end{equation}

The saddle-point equations are obtained by variations with respects to the eigenvalues $m_i$ and $u_a$:
\bea
	\label{SaddlePointEquationU1}-\frac{4N}{\lambda}m_i+\sum_{j\neq i}\frac{d}{dm_i}\log\mathcal{Z}_{1-\mathrm{loop}}(m_i-m_j)-\sum_{a}\frac{1}{e^{u_a-m_i}-1}=0,& \\
	\label{SaddlePointEquationU2}n+\sum_{b\neq a}\coth\frac{u_a-u_b}{2}+\sum_i\frac{1}{e^{u_a-m_i}-1}=0.&
\eea
To proceed, we make the following Ansatz \cite{Okuda:2007kh}. The eigenvalues $m_i$ are divided into two groups $\{m_i^{(1)}:i=1,\dots,n\}$ and $\{m_i^{(2)}:i=n+1,\dots,N\}$ separated by $u_a$, which are uniformly distributed along $[u,u+2\pi i]$ ($\max\{m_i^{(2)}\}\ll u\ll \min\{m_i^{(1)}\}$). As we will further see, the position of $u$ is related to various parameters in the system.

Let us now explain the limit in which the above Ansatz for the eigenvalues is verified. 
Notice that approximating 
\begin{equation}
	\frac{1}{e^x-1}\simeq -\Theta(-x),
\label{eq:stepfn}
\end{equation}
where $\Theta$ is Heaviside step function, and using that $\frac{d}{dx}\log\mathcal{Z}_{1-\mathrm{loop}}(x)\to0$ when $x\to\infty$, the saddle-point equations decouple and become 
\begin{equation}
\begin{split}
	-\frac{4N}{\lambda}m_i^{(1)}+\sum_{j\neq i}\frac{d}{dm_i^{(1)}}\log\mathcal{Z}_{1-\mathrm{loop}}(m_i^{(1)}-m_j^{(1)})+k=0,& \\
	-\frac{4N}{\lambda}m_i^{(2)}+\sum_{j\neq i}\frac{d}{dm_i^{(2)}}\log\mathcal{Z}_{1-\mathrm{loop}}(m_i^{(2)}-m_j^{(2)})=0,& \\
	n+\sum_{b\neq a}\coth\frac{u_a-u_b}{2}-n=0.&
\end{split}
\end{equation}
The last equation is satisfied when $u_a$ are uniformly distributed. The first two equations describe the same distribution of eigenvalues as in the partition function, centered at $k\lambda/4N$ and $0$, respectively. The only difference is that $\lambda$ needs to be rescaled as $\lambda\mapsto\lambda(\mathrm{\#~of~eigenvalues}/N)$, namely, $\lambda n/N$ and $\lambda (N-n)/N$, respectively. Finally, this solution for the eigenvalues is valid when the Heaviside approximation is correct which in turn means that the distance between the centers of the distributions, which is $k\lambda/4N$, should be large.

At this point, a couple of comments are in order.  Firstly, the requirement $k\lambda/4N\gg1$ not only ensures that the approximation (\ref{eq:stepfn}) is valid, but also guarantees that the derivative of $\mathcal{Z}_{1-\mathrm{loop}}(x)$ is small.  However, while the Heaviside approximation receives exponentially small corrections, the latter will receive power-law corrections.  Thus the above Ansatz will be power-law corrected.  Secondly, since we work in the large-$N$ limit, the number of columns of the rectangular representation, $k$, has to scale with $N$ in order to satisfy $k\lambda/4N\gg1$ for fixed $\lambda$.

Under this approximation, the expectation (\ref{WLwithUMatrix}) splits into three components
\begin{equation}
	W=W_1+W_2+W_3,
\end{equation}
where
\begin{align}
\label{eq:W1}
	W_1&=-\frac{2N}{\lambda} \sum_i \left(m_i^{(1)}\right)^2 +\sum_{i<j}\log \mathcal{Z}_{1-\mathrm{loop}}(m_i^{(1)}-m_j^{(1)})+k\sum_i m_i^{(1)},\\
	W_2&=-\frac{2N}{\lambda} \sum_i \left(m_i^{(2)}\right)^2 +\sum_{i<j}\log \mathcal{Z}_{1-\mathrm{loop}}(m_i^{(2)}-m_j^{(2)}),\\
	W_3&=\sum_{a<b}\log\left(2\sinh\frac{u_a-u_b}{2}\right)^2.
\end{align}
The first two terms are simply rescaled partition functions.  For $W_1$, we complete the square to remove the last term in (\ref{eq:W1}), while $W_2$ is already centered at the origin.  We thus have
\begin{align}
	W_1&=F\left(\lambda\frac{n}{N},n\right)+\frac{k^2 \lambda n}{8N},\nn\\
	W_2&=F\left(\lambda\left(1-\frac{n}{N}\right),N-n\right),
\end{align}
where $F(\lambda,N)$ is the effective free energy
\begin{equation}
	F(\lambda,N)=\log \mathcal{Z}(\lambda,N)-\log N!.
\end{equation}
The reason we subtract $\log N!$ is because the permutation symmetry in $m_i$ is not properly accounted for if we directly use the free energy, where the permutation symmetry will contribute $\log\left(N-n\right)!+\log n!-\log N!$ instead of $0$. Therefore, in order to obtain the correct result we need to remove this symmetry in the free energy. Note also that it is valid to include the one-loop correction in the free energy.

Finally, $W_3$ can be evaluated exactly when the $\{u_a\}$ are uniformly distributed
\begin{equation}
\begin{split}
	W_3&=\log(\pm1)+k(k-1)\log2+2\log\left(\prod_{a<b}\sin\frac{\pi(a-b)}{k}\right)\\
	&=\log(\pm1)+k(k-1)\log2+2\log\left(\prod_{i=1}^{k-1}\sin^\frac{k}{2}\left(\frac{\pi i}{k}\right)\right) \\
	&=\log(\pm1)+k(k-1)\log2+k\log\frac{k}{2^{k-1}} \\
	&=\log(\pm1)+k\log k.
\end{split}
\end{equation}

Therefore the vev of Wilson loop, (\ref{WLwithUMatrix}), is
\begin{equation}\label{WLOkuda}
	\left<W_\mathcal{R}\right>=\exp\left(\frac{k^2 \lambda n}{8N}+k\log k-k\log2\pi+F\left(\lambda\frac{n}{N},n\right)+F\left(\lambda\left(1-\frac{n}{N}\right),N-n\right)-F(\lambda,N)\right),
\end{equation}
where the factor $1/k!$ in (\ref{WLwithUMatrix}) is cancelled by the permutation symmetry of $u_a$. The term $k\log k$ will be cancelled by the one-loop correction, which will be discussed in detail in Section \ref{Sec:OneLoopCorrections} and the rest of the terms are at most $O(N^2\log \lambda)$, which  is subdominant to the leading term in the strong coupling limit $\lambda\to\infty$. Therefore the leading term is universal, and in particular is independent of the $\mathcal N=2^*$ hypermultiplet mass $M_0$, at least in this limit.

\subsection{Error estimation and numerical explorations}
Our computation of the vacuum expectation value of the rectangular Wilson loop relies on two key approximations. First, we assume that the leading order answer is given by a saddle point evaluation which requires the large-$N$ limit. Second, we assume that the characteristic eigenvalue distribution is given by two groups of eigenvalues which are widely separated with separation $k \lambda/4N$. Thus, for large separation we require large $k$ or large $\lambda$. We will provide a systematic way of introducing corrections to this leading order approximation in various parameters. As can be seen for the structure of the parameter, we need to consider corrections in $k$ and in $\lambda$.

Let us discuss the sources of error in the approximations made in the Ansatz above for the eigenvalue distribution. One important approximation relies on the fact that  $\frac{d}{dm_i^{(1)}}\log \mathcal{Z}_{1\mathrm{-loop}}(m_i^{(1)}-m_j^{(2)})$ changes slowly when $j$ runs over all possible values in the second group of eigenvalues.  In the strong-coupling limit (large $\lambda$), the contribution of such crossing terms is bounded by
\begin{equation}
	\sum \frac{2(1+M^2)}{\left(m^{(1)}_i-m^{(2)}_j\right)^2}\sqrt{\lambda(1-n/N)(1+M_0^2)}={\cal O}\left(\left(\frac{1+M_0^2}{\lambda}\right)^{3/2}\frac{N^3}{k^2}\right).
\end{equation}
Therefore when $\lambda\gg1+M_0^2$ and $k={\cal O}(N^{3/2})$, the above Ansatz is reliable. For symmetric Wilson loops, we need to carefully analyze the crossing terms, which will be performed in the next section.

\begin{figure}[t]
	\centering
	\includegraphics[scale=1.0]{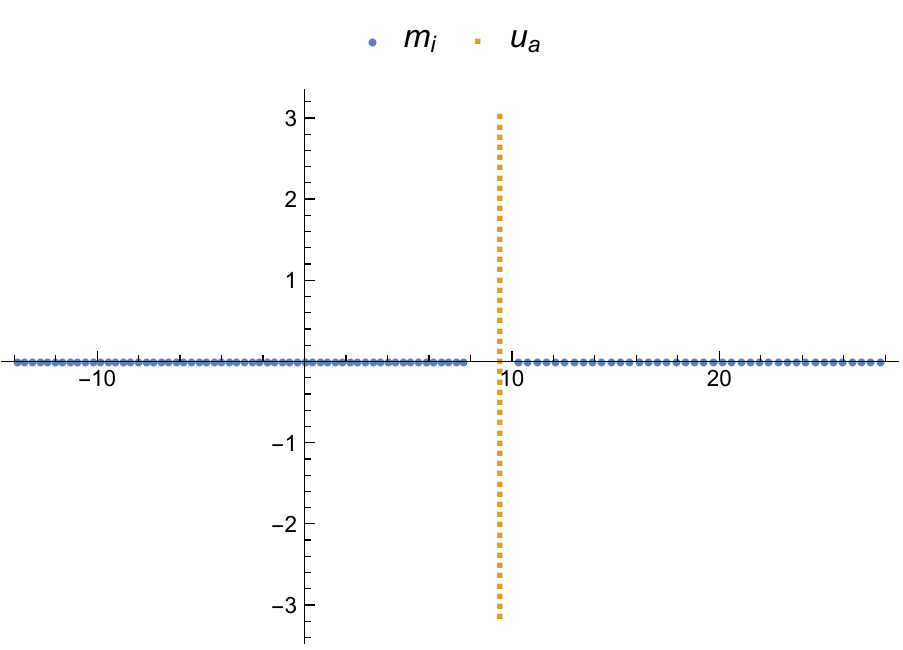}
\caption{Eigenvalue distribution in $\mathcal{N}=2^*$ theory. The parameters used in the plot above are 
$N = 100$, $\lambda = 100$, $k = 50$, $n=40$ and $M_0=2$.} 
\label{Fig:k50}
\end{figure}

\begin{figure}[t]
	\centering
	\includegraphics[scale=1.0]{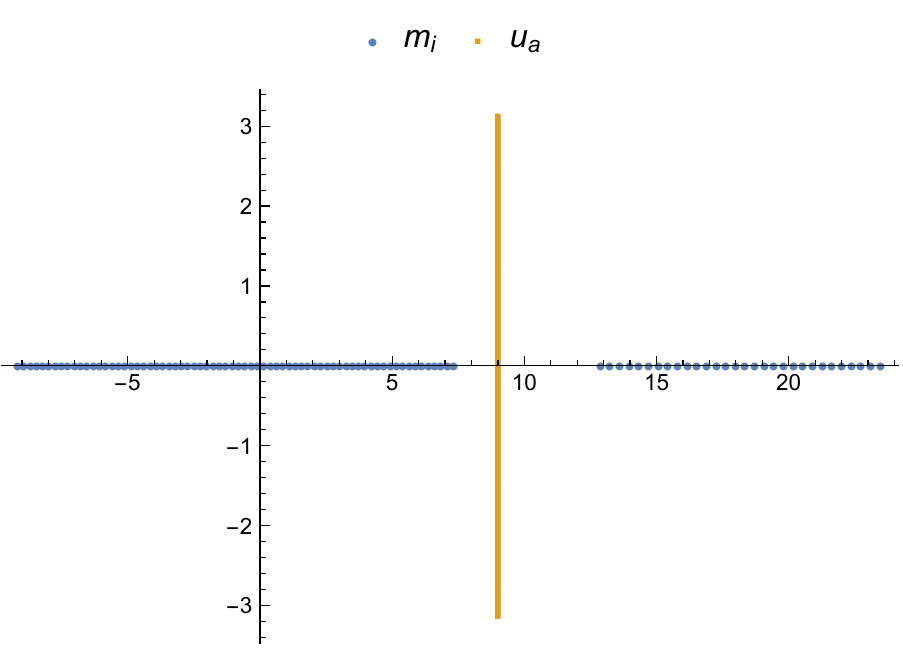}
\caption{Eigenvalue distribution in $\mathcal{N}=2^*$ theory. The parameters used in the plot above are $N=100$, $\lambda=20$, $k=300$, $n=30$ and $M_0=2$.}
\label{Fig:k300}
\end{figure}

In order to demonstrate more concretely the limits that we have used, we present numerical results for the eigenvalues for two choices of parameters. Figure~\ref{Fig:k50} represent the numerical solution to the saddle point equations for $N = 100$, $\lambda = 100$, $k = 50$ and $n=40$ and $M_0=2$ while Figure~\ref{Fig:k300} describes the distribution for $N=100$, $\lambda=20$, $k=300$, $n=30$ and $M_0=2$. 
The eigenvalue distribution in both cases, despite the wide range of the parameters, is qualitatively the same. It can also be explicitly seen from the plots that increasing $k$ makes the approximation better.

\section{Symmetric and Antisymmetric representations}\label{Sec:SandA}

In this section we discuss in detail the approximations that take place when $(n,k)$ above take values corresponding to the totally symmetric and totally antisymmetric representations. We recover well-established results in the context of ${\cal N}=4$ and those presented for ${\cal N}=2^*$ albeit in a rather subbtle fashion that sheds lights on various approximations implicitly made in the literature.

\subsection{Symmetric representation}
Let us start by comparing our result with the standard expression for the totally symmetric Wilson loop in ${\cal N}=4$ SYM that was obtained, for the first time, in  \cite{Hartnoll:2006is}:
\be
\label{Eq:HKSk}
\left<W_{S_k}\right>=\exp\left(2N\left(\kappa\sqrt{1+\kappa^2}+\sinh^{-1}\kappa\right)\right),
\ee
where $\kappa=k\sqrt{\lambda}/4N$. Our general expression, given in equation (\ref{WLOkuda}), is our starting point. First, we particularize to the totally symmetric representation: $k=fN, n=1$. Since we assumed large $\lambda$ in Eq. (\ref{WLOkuda}) we impose such approximation in Eq. (\ref{Eq:HKSk}), namely, for large  $\lambda$, we approximate  $\sqrt{1+\kappa^2}\sim\kappa$ and $\sinh^{-1}\kappa<<\kappa$. In this approximation the leading term in both, Eq. (\ref{Eq:HKSk}) and Eq. (\ref{WLOkuda}) coincide
\be
\left<W_{S_k}\right>\simeq\exp\left(2N\kappa^2\right)=\exp\left(\frac{f^2\lambda N}{8}\right)=\exp\left(\frac{k^2\lambda n}{8N}\right).
\ee

It is now clear that to improve in the comparison to Eq. (\ref{Eq:HKSk}) we need to consider terms subleading in $\lambda$. Going back to the saddle point configuration of eigenvalues, we recall that  the interaction between $m_i^{(1)}$ and $m_i^{(2)}$ was ignored assuming that the distance between them is very large ($O(k\lambda/4N)$). However, in our current situation, the eigenvalue $m\equiv m_1^{(1)}$ is affected by $(N-1)$ $m_i^{(2)}$, so the interaction is of order $O(N(1+M_0^2)/f\lambda)$, which is not negligible and hence the position of $m$ should be corrected accordingly. The saddle-point equation for $m$ containing all the other eigenvalues is
\be\label{mEquation}
-\frac{4N}{\lambda}m+\sum_{j>1}\frac{2(1+M_0^2)}{m-m_j}+k=0.
\ee
It is obvious that (\ref{mEquation}) is the same equation as in $\mathcal{N}=4$ theory, except that $\lambda$ and $f$ need to be rescaled accordingly ($\lambda\mapsto\lambda(1+M_0^2),f\mapsto f/(1+M_0^2)$).

In the large $N$ limit, the summation can be replaced by an integral
\begin{equation}
	\sum_{j>1}\frac{2(1+M_0^2)}{m-m_j}\simeq (N-1)\int_{-\sqrt{\lambda_r^*}}^{\sqrt{\lambda_r^*}}dx\rho(x)\frac{2(1+M_0^2)}{m-x}=\frac{4N}{\lambda}\left(m-\sqrt{m^2-\lambda_r^*}\right).
\end{equation}
where $\lambda^*=(1-1/N)\lambda_r$ and $\lambda_r=(1+M_0^2)\lambda$. For convenience we also define $\kappa_r=\kappa/\sqrt{1+M_0^2}$. Then (\ref{mEquation}) gives the position of $m$
\begin{equation}\label{mPosition}
	m=\sqrt{\lambda_r^*+\kappa_r^2 \lambda_r}.
\end{equation}

The crossing terms in saddle-point equation for the other $m_i$ is
\begin{equation}
	\frac{2(1+M_0^2)}{m}=O\left(\frac{1+M_0^2}{\lambda}\right),
\end{equation}
which can be safely ignored in the strong-coupling limit.

Therefore the vev of Wilson loop is obtained by evaluating
\begin{equation}\label{FullWLSk}
\begin{split}
	\left<W_{S_k}\right>=\exp\left[-\frac{2N}{\lambda}m^2+km+2(1+M_0^2)\sum_{i>1}\log(m-m_i)+k\log k-\right. \\
	\left.-k\log 2\pi+F\left(\lambda_r\left(1-\frac{1}{N}\right),N-1\right)-F(\lambda_r,N)\right]
\end{split}
\end{equation}
at the saddle-point.

The first two terms simply give
\begin{equation}\label{KineticTerms}
	-\frac{2N}{\lambda}m^2+km=2N(1+M_0^2)\left[2\kappa_r\sqrt{\kappa_r^2+\lambda_r^*/\lambda_r}-(\kappa_r^2+\lambda_r^*/\lambda_r)\right],
\end{equation}
and the summation over $i$ can be replaced by an integral
\begin{equation}\label{LogTerms}
\begin{split}
	&2(1+M_0^2)\sum_{i>1}\log(m-m_i) \\
	\simeq&2(N-1)(1+M_0^2)\int_{-\sqrt{\lambda_r^*}}^{\sqrt{\lambda_r^*}}dx\rho(x)\log(m-x) \\
	=&2(N-1)(1+M_0^2)\left[\log m-\frac{\lambda_r^*}{8m^2}\tensor[_3]{F}{_2}\left(1,1,\frac{3}{2};2,3;\frac{\lambda_r^*}{m^2}\right)\right] \\
	=&2(N-1)(1+M_0^2)\left[\log\left(\kappa_r+\sqrt{\kappa_r^2+\frac{\lambda_r^*}{\lambda_r}}\right)+\frac{1}{2}\log \lambda_r-\log 2-\frac{1}{2}+\left(1+\frac{\lambda_r}{\lambda_r^*}\kappa_r^2\right)-\frac{\lambda_r}{\lambda_r^*}\kappa_r\sqrt{\kappa_r^2+\frac{\lambda_r^*}{\lambda_r}}\right]
\end{split}
\end{equation}
where $\tensor[_p]{F}{_q}(a_1,\dots,a_p;b_1,\dots,b_q;z)$ is the generalized hypergeometric function.

The difference of free energy is (see (\ref{EffectiveFreeEnergy}) for more detail)
\begin{equation}\label{DifferenceOfFreeEnergy}
	F\left(\lambda_r^*,N-1\right)-F(\lambda_r,N)=N(-\log\lambda_r+1+2\log 2)-\log2-\log 2\pi+\frac{1}{2}\log\lambda_r.
\end{equation}

At the leading order, $\lambda_r^*\simeq\lambda_r$ and $N-1\simeq N$, so the first term in (\ref{LogTerms}) can be written as a arcsinh function, the last two terms is cancelled by the corresponding terms in (\ref{KineticTerms}). Besides, the $k\log k$ in (\ref{FullWLSk}) will be cancelled by the one-loop correction. Therefore, at leading order the above expression can be simplified to
\begin{equation}\label{WLSk}
	\left<W_{S_k}\right>_{\mathrm{planar}}=\exp\left(2N(1+M_0^2)G\left(\frac{\kappa}{\sqrt{1+M_0^2}}\right)\right),
\end{equation}
where
\begin{equation}
	G(x)=x\sqrt{1+x^2}+\sinh^{-1}(x)-\frac{f}{1+M_0^2}\log 2\pi.
\end{equation}
Our result (\ref{WLSk}) matches with (2.6) in \cite{Chen-Lin:2015xlh} up to a $O(f)$ term, which will be cancelled after considering the one-loop correction and the validity of saddle-point approximation. Now we have explicitly shown  that by including the interactions between the two sets of eigenvalues we obtain a more precise expression for the expectation value of the Wilson loop.

Up until now the rectangular Wilson loop has been  evaluated with help of (\ref{2UMatrix}); we have shown that various higher rank Wilson loops are thus evaluated in a unifying framework.  However, for the totally symmetric representation, the identity (\ref{2VMatrix}) could also be employed. Although both pictures are equivalent, it is clear that the role the parameters enters is different, thus providing an alternative approximation. 

Similar to (\ref{WLwithUMatrix}), the Wilson loop can be written as
\begin{equation}
\begin{split}
	\left<W_\mathcal{R}\right>=\frac{1}{n!(2\pi)^n}\frac{1}{\cal Z}\int\prod_{a=1}^n dv_a \prod_{i=1}^N dm_i \exp\left[-\frac{2N}{\lambda} \sum_{i=1}^N m_i^2 +\sum_{i<j}\log \mathcal{Z}_{1-\mathrm{loop}}(m_i-m_j)+k\sum_{a=1}^n v_a\right. \\
	 \left.+\sum_{a<b}\log\left(2\sinh\frac{v_a-v_b}{2}\right)^2-\sum_{a,i}\log(1-e^{m_i-v_a})\right],
\end{split}
\end{equation}
and the saddle-point equation is
\bea
-\frac{4N}{\lambda}m_i+\sum_{j\neq i}\frac{d}{dm_i}\log\mathcal{Z}_{1-\mathrm{loop}}(m_i-m_j)+\sum_{a}\frac{1}{e^{v_a-m_i}-1}=0,& \\
	k+\sum_{b\neq a}\coth\frac{v_a-v_b}{2}-\sum_i\frac{1}{e^{v_a-m_i}-1}=0.&
\eea
Compared with (\ref{SaddlePointEquationU1}) and (\ref{SaddlePointEquationU2}), the force between eigenvalues $v_a$ and $m_i$ becomes attractive. For the symmetric representation ($k=fN,n=1$), we make the following ansatz: $m_1$ is located far away from the the rest $N-1$ eigenvalues, with the eigenvalues $m_i$'s  centered at $0$, and $v$ is very close to $m_1$. In the strong-coupling limit $\lambda\to\infty$, the saddle-point equations are greatly simplified
\bea
-\frac{4N}{\lambda}m_1+\sum_{j\neq i}\frac{2(1+M_0^2)}{m_1-m_j}+\frac{1}{e^{v-m_1}-1}=0,& \\
-\frac{4N}{\lambda}m_i+\sum_{j\neq i,j>1}\frac{2(1+M_0^2)}{m_i-m_j}=0,& \\
	k-\frac{1}{e^{v-m_1}-1}=0.&
\eea

From the first and the last equation  we get exactly the same equation as (\ref{mEquation}). Therefore, the position of eigenvalues $m_1$ is precisely the same as (\ref{mPosition}) in both approaches and at the leading order the vev of Wilson loop is clearly the same as (\ref{WLSk}).


\subsection{Anti-symmetric representation}

For the totally anti-symmetric representation ($n=fN$ and $k=1$), the previous saddle-point configuration is not valid since $\frac{k\lambda}{4N}$ is small and the interaction between the two groups of eigenvalues cannot be ignored. However, from our intuition in the general rectangular case when we decrease $k$, the saddle-point configuration is supposed to change continuously; besides, inserting only one eigenvalue $u_1$ will not considerably distort the Wigner distribution. We, therefore, suggest the following saddle-point configuration: the eigenvalues $m_i$ obey the Wigner distribution and $u_1$ is inserted into an equilibrium position, i.e.
\begin{equation}
	n-\sum_{i=1}^{N}\frac{1}{e^{u_1-m_i}-1}=0.
\end{equation}
Approximating  the sum by an integral, we get
\begin{equation}
	f-\int dx \frac{\rho(x)}{e^{u-x}-1}=0,
\end{equation}
whose solution is $-z_0$ in the strong-coupling limit $\lambda>>1$, where $z=z_0/\sqrt{\lambda}$ is the saddle-point in \cite{Hartnoll:2006is}
\begin{equation}\label{AntiSymmetricSaddlePoint}
	\cos^{-1}z-z\sqrt{1-z^2}=\pi(1-f).
\end{equation}
Since the anti-symmetric representation has the symmetry $f\mapsto 1-f,z\mapsto -z$, our result agrees completely with \cite{Hartnoll:2006is}
\begin{equation}
	\left<W_{A_n}\right>=\exp\left(\frac{2N}{3\pi}\sqrt{\lambda}\sin^3\theta_k\right).
\end{equation}

Now, consider the rectangular representation $n=fN, k={\cal O}(1)$, we suggest the following saddle-point configuration: the position of $m_i$ does not change and $u_a$ are uniformly distributed along $[-z_0,-z_0+2\pi i]$. The saddle-point equation for $u_a$ is
\begin{equation}
	n+\sum_{b\neq a}\coth\frac{u_a-u_b}{2}+\sum_{i=1}^{N}\frac{1}{e^{u_a-m_i}-1}=0,
\end{equation}
where the second term vanishes and the last term is independent of the imaginary part of $u_a$ when $\lambda$ is large, since it will be approximated by a step function. Therefore at leading order the vev of Wilson loop is just
\begin{equation}
	\left<W_\mathcal{R}\right>=k!\left<W_{A_n}\right>^k.
\end{equation}

\section{One loop corrections}\label{Sec:OneLoopCorrections}

Going beyond the leading term is an important first step into the intricate structure of Wilson loops. The classical example is the computation of the L\"uscher term in confining theories which is determined by the number of light degrees of freedom on the effective chromoelectric flux tube \cite{Luscher:1980ac}. In supersymmetric theories and in the context of the AdS/CFT correspondence there is the added interest in this corrections as they can be obtained, in principle, using the dual gravity theory. 

In this section we closely follow the paradigm of \cite{Ambjorn:1992gw} (see also the pedagogical exposition of \cite{Eynard:2015aea}). The main result of those work is the construction of a systematic way to compute higher order corrections. In appendix \ref{App:genus} we briefly stated the main elements of the construction and applied to reproduced the first sub-leading term in the Wilson loop in the fundamental representation.

The calculation for one-loop correction is an application of multi-dimensional steepest descent formula
\begin{equation}
	\int g(x)\exp(N f(x))d^dx=\frac{1}{N^{d/2}}\exp(Nf(c))I\left(\frac{1}{N}\right),
\end{equation}
where $f$ has global maximum at a unique point $c$ (the Hessian matrix $A_{ij}\equiv\partial^2 f/\partial x_i\partial x_j$ is negative definite) and $I(x)$ extends to a smooth function on $[0,\infty)$ such that
\begin{equation}
	I(0)=(2\pi)^\frac{d}{2}\frac{g(c)}{\sqrt{\det A}}.
\end{equation}

Therefore in order to calculate the one-loop correction for Wilson loop, we need to evaluate the determinant of the Hessian matrix $\det A$, collect the factors $N^{-d/2}$ and $(2\pi)^{d/2}$, and then add them together to obtain the one-loop correction
\begin{equation}
	-\frac{1}{2}\log\det A+\frac{d}{2}(\log 2\pi-\log N)=-\frac{1}{2}\log\det (NA)+\frac{d}{2}\log 2\pi.
\end{equation}

\subsection{One-loop correction for rectangular representation}\label{Subsec:Hessian}

There are $N+k$ variables in the integrand, so the contribution of one-loop determinant is
\begin{equation}\label{LogOneLOopDeterminant}
	W_{\mathrm{det}}=-\frac{1}{2}\log\det_{x,x^\prime} A_{x,x^\prime}+\frac{k}{2}\log2\pi+\frac{1}{2}\log\det_{y,y^\prime} B_{y,y^\prime},
\end{equation}
where the Hessian matrices $A_{x,x^\prime}$ and $B_{y,y^\prime}$ ($x,x^\prime$ run over all the variables $m_i$ and $u_a$ while $y,y^\prime$ just run over $m_i$) are given by
\begin{equation}
A_{x,y}=\left(
\begin{array}{ccc}
 A_{i,j}^{(1)} & -\left(\log  \mathcal{Z}_{1-\text{loop}}\right)''\left(m_i^{(1)}-m_j^{(2)}\right) & \frac{e^{-m_i^{(1)}+u_a}}{\left(-1+e^{-m_i^{(1)}+u_a}\right)^2}
\\
 -\left(\log  \mathcal{Z}_{1-\text{loop}}\right)''\left(m_i^{(2)}-m_j^{(1)}\right) & A_{i,j}^{(2)} & \frac{e^{-m_i^{(2)}+u_a}}{\left(-1+e^{-m_i^{(2)}+u_a}\right)^2}
\\
 \frac{e^{-m_i^{(1)}+u_a}}{\left(-1+e^{-m_i^{(1)}+u_a}\right)^2} & \frac{e^{-m_i^{(2)}+u_a}}{\left(-1+e^{-m_i^{(2)}+u_a}\right)^2} & A_{a,b}
\\
\end{array}
\right)
\end{equation}

\begin{equation}
A_{i,i}^{(q)}=-\frac{4N}{\lambda }+\sum _{(j,r)\neq (i,q)} \left(\log  \mathcal{Z}_{1-\text{loop}}\right)''\left(m_i^{(q)}-m_j^{(r)}\right)-\sum _{a=1}^k
\frac{e^{-m_i^{(q)}+u_a}}{\left(-1+e^{-m_i^{(q)}+u_a}\right)^2},q=1,2
\end{equation}

\begin{equation}
A_{i,j}^{(q)}=-\left(\log  \mathcal{Z}_{1-\text{loop}}\right)''\left(m_i^{(q)}-m_j^{(q)}\right)^2,i\neq j
\end{equation}

\begin{equation}
A_{a,a}=-\frac{1}{2}\sum _{b\neq a} \text{csch}^2\left(\frac{u_a-u_b}{2}\right)-\sum _{i=1}^N \frac{e^{-m_i+u_a}}{\left(-1+e^{-m_i+u_a}\right)^2}
\end{equation}

\begin{equation}
A_{a,b}=\frac{1}{2}\sum _{b\neq a} \text{csch}^2\left(\frac{u_a-u_b}{2}\right),a\neq b
\end{equation}

\begin{equation}
	B_{m_i,m_i}=-\frac{4N}{\lambda}+\sum _{j\neq i} \left(\log  \mathcal{Z}_{1-\text{loop}}\right)''(m_i-m_j)
\end{equation}

\begin{equation}
	B_{m_i,m_j}=-\left(\log  \mathcal{Z}_{1-\text{loop}}\right)''(m_i-m_j)^2
\end{equation}

Under the approximation \(\left| m_i^{(1)}-m_j^{(2)}\right| >>1\) (i.e. our previous saddle-point configuration is valid), \((e^{u_a-m_i}-1)^{-1}\)
can be approximated by a step function and all the off-diagonal terms vanish. It can be checked that 

\begin{equation}
A_{x,y}\to \left(
\begin{array}{ccc}
 A_{i,j}^{(1)} & 0 & 0 \\
 0 & A_{i,j}^{(2)} & 0 \\
 0 & 0 & A_{a,b} \\
\end{array}
\right)
\end{equation}

\begin{equation}
A_{a,b}=-\frac{1}{2}\frac{k^2-1}{3} \delta_{a,b}+\frac{1}{2}\tilde{A}_{a,b}
\end{equation}

\begin{equation}
\tilde{A}_{a,b}=
\left\{
\begin{array}{ll}
 \sum _{b\neq a} \csc ^2\left(\pi (a-b)/k\right) & 1\leq a\neq b\leq k \\
 0 & \text{otherwise} \\
\end{array}
\right.
\end{equation}

The eigenvalues of \(\tilde{A}\) is
\begin{equation}
\frac{1}{3}\left(k^2-1\right)-2n(k-n),\qquad n=0,1,\ldots ,k-1.
\end{equation}

Notice that there is a zero mode in $A_{a,b}$, which is due to the translational symmetry in the imaginary direction, we should remove the zero mode and multiply the result by $2\pi$. The determinant of \(A_{a,b}\) (after removing the zero mode) is
\begin{equation}
{\det}^\prime A_{ab}=\prod_{n=1}^{k-1}n(k-n)=[(k-1)!]^2
\end{equation}
so the one-loop correction coming from \(u_a\) is given by
\begin{equation}\label{OneLoopDeterminantU}
-\frac{1}{2}\log  \det  A_{a,b}=-k\log k+k-\frac{1}{2}\log2\pi+\frac{1}{2}\log k
\end{equation}

The contribution coming from $A_{i,j}^{(q)}$ is contained in the exact formula of free energy. For Gaussian matrix model
\begin{equation}
	\mathcal{Z}=(2\pi)^{N/2}\left(\frac{4N}{\lambda}\right)^{-N^2/2}G(1+N)N!
\end{equation}
where $G$ is the Barnes-G function, so the effective free energy is
\begin{equation}\label{EffectiveFreeEnergy}
\begin{split}
	F(\lambda,N)=\log\mathcal{Z}-\log N!\simeq N^2\left(\frac{1}{2}\log\lambda-\frac{3}{4}-\log 2\right)+N\log 2\pi-\frac{1}{12}\log N+\frac{1}{12}-\log A
\end{split}
\end{equation}
where $A$ is the Glaisher-Kinkelin constant.  Collecting everything together, we find
\begin{equation}
	\left<W_\mathcal{R}\right>=\exp\left(\frac{k^2 n \lambda}{8N}+a_0 N^2+a_1 N \log N + a_2 N + a_3 \log N + a_4\right),
\end{equation}
where $f\equiv k/N,g\equiv n/N\leq 1$ and
\eqa
&a_0 = (g^2-g)\log\lambda+g(1-g)\left(2\log 2+\frac{3}{2}\right)+\frac{1}{2}\left[(1-g)^2\log(1-g)+g^2\log g\right]\\
&a_1 = 0\\
&a_2 = f-\frac{1}{2}f\log(2\pi)\\
&a_3 = \frac{5}{12}\\
&a_4 = \frac{1}{12}+\frac{\log f}{2}-\frac{1}{12}\log g(1-g)-\log A
\eea

The symmetry $n\mapsto N-n$ ($g\mapsto 1-g$) is preserved at the one-loop level. It is also obvious that $a_0\sim\log \lambda$, so when $k$ is large enough ($k^2n={\cal O}(N^3)$), the rectangular Wilson loop is dominated by the leading term $k^2 n\lambda/8N$, which justifies our previous statement.

\subsection{One-loop correction for the totally symmetric representation}

Now, let us focus on the $k$-symmetric representation (\(n=1\)). For simplicity we consider the \(\mathcal{N}=4\) theory (for \(\mathcal{N}=2^*\)
theory we only need to re-scale \(f\) and \(\lambda\))

The symmetric representation is just a special case of rectangular representation ($n=1$). Notice that the one-loop correction coming from \(A^{(1)}\) is given by
\begin{equation}
\frac{1}{2}\log 2\pi-\frac{1}{2}\log  \left[\frac{4N}{\lambda }+\sum _{j>1} \frac{2}{\left(m-m_j\right)^2}\right]\simeq \frac{1}{2}\log2\pi-\log  2-\frac{1}{2}\log  N+\frac{1}{2}\log\frac{\lambda\kappa}{\sqrt{1+\kappa^2}},
\end{equation}
the correction coming from $u_a$ is (\ref{OneLoopDeterminantU})
and correction coming from the difference of free energy (\ref{DifferenceOfFreeEnergy}) is
\begin{equation}
	-\log2-\log 2\pi+\frac{1}{2}\log\lambda,
\end{equation}
the contribution of whole one-loop determinant (\ref{LogOneLOopDeterminant}) is
\begin{equation}
	W_{\mathrm{det}}=\frac{k-1}{2}\log 2\pi-k\log k+k+\frac{1}{2}\log f-2\log2+\frac{1}{2}\log\frac{\kappa}{\sqrt{1+\kappa^2}}+\log\lambda
\end{equation}
and at the one-loop level is
\begin{equation}
	W_{\mathrm{det}}^{1\mathrm{-loop}}=-\frac{1}{2}\log 2\pi+\frac{1}{2}\log f-2\log2+\frac{1}{2}\log\frac{\kappa}{\sqrt{1+\kappa^2}}+\log\lambda.
\end{equation}
Notice that the logarithm of one-loop determinant (\ref{OneLoopDeterminantU}) contains a term $W_{\mathrm{det}}-W_{\mathrm{det}}^{1\mathrm{-loop}}=k+\frac{1}{2}k\log2\pi=O(N)$, it actually corrects the leading order behavior of symmetric Wilson loops.

We also need to keep track of the difference between $\lambda^*\equiv\lambda(1-1/N)$, $N-1$ and $\lambda$, $N$ carefully.
Expand (\ref{KineticTerms}) and (\ref{LogTerms}) to the sub-leading order w.r.t. $1/N$, we get the correction coming from the interaction between $m$ and $m_i$
\begin{equation}
	W_{\mathrm{int}}=-2\log(\kappa+\sqrt{1+\kappa^2})-\log \lambda+2\log 2+1-\frac{\kappa}{\sqrt{1+\kappa^2}}.
\end{equation}

The one-loop corrected symmetric Wilson loop is
\begin{equation}
	\left< W_{S_k}\right>_{1\mathrm{-loop}}=\left<W_{S_k}\right>_{\mathrm{planar}}+W_{\mathrm{det}}+W_{\mathrm{int}}.
\end{equation}
or more explicitly
\begin{equation}\label{UMatrix1LoopSymmetricWL}
	\left< W_{S_k}\right>_{1\mathrm{-loop}}\simeq 2N\left(\kappa \sqrt{1+\kappa ^2}+\sinh^{-1}\kappa+\frac{f}{2}-\frac{f}{4}\log 2\pi\right)-\frac{1}{2}\log\frac{2\pi (\kappa+\sqrt{1+\kappa^2})^4}{f\kappa/\sqrt{1+\kappa^2}}.
\end{equation}

The holographic calculation of one-loop corrected expectation value for the symmetric representation, using the spectrum of excitations previously obtained in \cite{faraggi:2011bb}, was presented in \cite{Buchbinder:2014nia}:
\begin{equation}
\label{Eq:holoSk}
\left\langle W_{S_k}\right\rangle =2N\left(\kappa \sqrt{1+\kappa ^2}+\sinh^{-1} \kappa \right)-\frac{1}{2}\log  \frac{\kappa ^3}{\sqrt{1+\kappa^2}}.
\end{equation}
compared to which we find an extra term
\begin{equation}
2N\left(\frac{f}{2}-\frac{f}{4}\log 2\pi\right)
\end{equation}
at leading order, which is due to the saddle-point approximation for the variable $u_a$. Consider the identity (\ref{2UMatrix})
\begin{equation}
	\Tr_{S_k} e^M=\frac{1}{(2\pi)^kk!}\int \left[\prod_{a=1}^k du_a \right]\exp\left[\sum_{a=1}^k u_a+\sum_{a<b}\log\left(2\sinh\frac{u_a-u_b}{2}\right)^2+\sum_{a,i}\log(1-e^{m_i-u_a})\right]
\end{equation}
evaluated at the saddle-point configuration $m_i$. The left hand side simply gives the symmetric Wilson loop operator
\begin{equation}\label{SymmetricWLOperator}
	\Tr_{S_k}e^M=\sum_{i_1\leq i_2\leq\dots\leq i_k}e^{m_{i_1}+m_{i_2}+\dots+m_{i_k}}\simeq e^{km},
\end{equation}
because all the other terms are exponentially suppressed when evaluated at the saddle-point $m>>m_i$. The saddle-point approximation can be applied to the right hand side, and the solution is the same as above: $u_a$ are uniformly distributed along the imaginary direction and are located between the two groups of eigenvalues. The one-loop determinant is also the same as above, and therefore the right hand side is
\begin{equation}
	\exp\left[km+2N\left(\frac{f}{2}-\frac{f}{4}\log2\pi\right)+\frac{1}{2}\log f+\frac{1}{2}\log N-\log 2\pi\right]
\end{equation}

It is expected to produce the same result as (\ref{SymmetricWLOperator}), so the difference implies that the saddle-point method is not very precise and hence the error need to be subtracted from (\ref{UMatrix1LoopSymmetricWL}). Therefore the corrected symmetric Wilson loop is given by
\begin{equation}\label{UResult}
	\left<W_{S_k}\right>=2N\left(\kappa \sqrt{1+\kappa ^2}+\sinh^{-1}\kappa\right)-\frac{1}{2}\log N-\frac{1}{2}\log\frac{(\kappa+\sqrt{1+\kappa^2})^4}{2\pi\kappa/\sqrt{1+\kappa^2}}.
\end{equation}
There are various interesting features of this expression if one compares it to the holographic computation in \cite{Buchbinder:2014nia}. First, notice that it differs from the previous one-loop corrected field theory result of  \cite{Faraggi:2014tna} in that we included not only the Hessian contribution but corrections due to interactions between eigenvalues.  In the regime of large separation, $k \lambda /(4N)$, between the two groups of eigenvalues that we work in,  it is reasonable to consider large $\kappa = k \sqrt{\lambda}/(4N)$. In this limit we find that  Eq. (\ref{UResult}) goes as $-\log\kappa^2$ which is twice the value of the corresponding limit in holographic expression quoted in Eq. (\ref{Eq:holoSk}).

We can alternatively use the $V$ matrix (\ref{2VMatrix}) to evaluate the one-loop correction. Now the one-loop correction comes from the determinant of such a sub-matrix (after using the exact formula of free energy)
\begin{equation}
\begin{split}\label{V1LoopDet}
	&\det\left(
	\begin{matrix}
		\frac{e^{m-v}}{1-e^{m-v}}+\frac{e^{2 m-2 v}}{\left(1-e^{m-v}\right)^2}-\left[\frac{4N}{\lambda }+\sum _{j>1} \frac{2}{\left(m-m_j\right)^2}\right] & -\frac{e^{m-v}}{1-e^{m-v}}-\frac{e^{2 m-2 v}}{\left(1-e^{m-v}\right)^2} \\
		-\frac{e^{m-v}}{1-e^{m-v}}-\frac{e^{2 m-2 v}}{\left(1-e^{m-v}\right)^2} & \frac{e^{m-v}}{1-e^{m-v}}+\frac{e^{2 m-2 v}}{\left(1-e^{m-v}\right)^2}+O(e^{-m})
	\end{matrix}
	\right) \\
	&=\det\left(\begin{matrix}
		k(k+1)-\frac{4N\sqrt{1+\kappa^2}}{\lambda\kappa} & -k(k+1) \\
		-k(k+1) & k(k+1)
	\end{matrix}
	\right) \\
	&=-k(k+1)\frac{4N\sqrt{1+\kappa^2}}{\lambda\kappa}.
\end{split}
\end{equation}
After collecting everything together we obtain
\begin{equation}
	\left<W_{S_k}\right>\simeq\exp\left[2N\left(\kappa\sqrt{1+\kappa^2}+\sinh^{-1}\kappa\right)-\frac{3}{2}\log N-\frac{1}{2}\log\frac{(2\pi)^2(\kappa+\sqrt{1+\kappa^2})^4f^2}{\kappa/\sqrt{1+\kappa^2}}\right].
\end{equation}
Notice that the $f,\lambda$-dependence in the one-loop correction have the correct sign, but the coefficient is different from the holographic calculation. Besides, there is also an extra $\log N$ term, but the coefficient is $-3/2$, where the extra $\log N$ comes from the coupling between $v$ and $m$ in one-loop determinant. Therefore the validity of saddle-point approximation of variable $v$ is skeptical. Notice that there does not exist a large variable coupled with $v$ ($kv$ will be cancelled by $\sum\log(1-e^{m_i-v})$), we do need to evaluate the integral over $v$ explicitly.

In \cite{Chen-Lin:2015dfa}, it is suggested that the contour of $v$ is deformed so that the contour integral picks up the pole at $m$ and the rest part is exponentially suppressed. Hence the symmetric Wilson loop is reduced to the following integral
\begin{equation}
	\left<W_{S_k}\right>=\frac{1}{\cal Z}\int \prod_{i=1}^N dm_i \exp\left[-\frac{2N}{\lambda} \sum_{i=1}^N m_i^2 +\sum_{i<j}\log \mathcal{Z}_{1-\mathrm{loop}}(m_i-m_j)+k m\right],
\end{equation}
and hence we need to remove the factor $k(k+1)$ from the one-loop determinant (\ref{V1LoopDet}). The final result is
\begin{equation}
	\left<W_{S_k}\right>\simeq\exp\left[2N\left(\kappa\sqrt{1+\kappa^2}+\sinh^{-1}\kappa\right)-\frac{1}{2}\log N-\frac{1}{2}\log\frac{(2\pi)(\kappa+\sqrt{1+\kappa^2})^4}{\kappa/\sqrt{1+\kappa^2}}\right],
\end{equation}
which is the same as (\ref{UResult}) up to a $\log 2\pi$.
\subsection{Comments on the cumulant expansion}
One important element of the  systematic $1/N$ expansion is the role of the cumulant expansion. Namely, we find that generically:
\be
\langle \exp(A)\rangle = \exp(\langle A\rangle +\left<A^3\right>_c+\dots),
\ee
where the first term is the leading order in $N$.

In  this section we  verify   the results of \cite{Faraggi:2014tna}. More generally, we demonstrate that the cumulant expansion does not affect the naive computation of corrections beyond the saddle point even for ${\cal N}=2^*$ theory. Namely, we show that in various situations there are no $1/N$ corrections other than the ones discussed in subsection \ref{Subsec:Hessian} and computed by the Hessian around the saddle point. Indeed, although we do not provide any proof here, we have checked that a similar statement can be formulated about higher rank Wilson loops in ABJM as computed in \cite{Cookmeyer:2016dln}. 

For $\mathcal{N}=4$ SYM, the cumulant expansion vanishes at sub-leading order, which can be derived from orthogonal polynomials (see appendix \ref{App:genus} for some technical details and a relevant example).  Not only the strong-coupling limit of $\mathcal{N}=2^*$ SYM shares the same property but also ABJM theory.

There is a more general argument for the vanishing of second order cumulant in most matrix models; here we take $\mathcal{N}=2^*$ SYM as an example. Let $A$ be a function of $m_i$ and $\rho(m_1,\dots,m_N)$ be the probability distribution of eigenvalues $m_i$. Then, when  one integrates out $N-2$ eigenvalues, the resulting two-point density can be evaluated by saddle-point approximation
\begin{equation}
\begin{split}
	\rho(m_1,m_2)&=\exp\left[-\frac{2N}{\lambda}(m_1^2+m_2^2)-F(\lambda(1-\frac{2}{N}),N-2)+\right.\\
	&\left.+\sum_{j=1}^2N\int dm\rho(m)\log\mathcal{Z}_{1-\mathrm{loop}}(m_j-m)+O(1)\right] \\
	&=\exp(N g(m_1)+Ng(m_2)+O(1))
\end{split}
\end{equation}  
where $g(m_j)$ is a function depending on the concrete form of $\mathcal{Z}_{1-\mathrm{loop}}$.

Therefore, the sub-leading order of cumulant expansion is
\begin{equation}
	\left<A^2\right>_c\propto\left(\int dx \rho(x_1,x_2)\rho(x_3,x_4)(A(x_1)A(x_2)-A(x_1)A(x_3))\right)
\end{equation}
which can be evaluated using saddle-point approximation again
\begin{equation}
	\left<A^2\right>_c\propto(A(x_1^*)A(x_2^*)-A(x_1^*)A(x_3^*))=0,
\end{equation}
where $x_i^*$ is the saddle-point $g(x^*)=0$. Since
\begin{equation}
	\left<A^k\right>_c=O(N^{2-k}),
\end{equation}
only the leading order is needed when evaluating  one-loop corrections \cite{Eynard:2005}.

\section{Conclusions}\label{Sec:Conclusions}

In this manuscript we have computed the vacuum expectation value for supersymmetric Wilson loops  in representations of $U(N)$ described by rectangular Young tableaux with $n$ rows and $k$ columns. We presented a number of analytical results and verified the robustness of our main approximation with a combination of numerical and analytical techniques. 

 Our more general point of view allows to better  understand the structure of corrections of Wilson loops in large representations. For example, we have clarified how certain results in the literature are obtained explicitly in the large-$N$ limit but contain other implicit assumptions. Our scheme relies on the approximation that the eigenvalues distribute into two groups. The key parameter in this framework is the distance between the two  groups of eigenvalues which is $k \lambda/ (4N)$. We gave a clear interpretation to the structure of corrections by carefully incorporating interactions among the two groups of eigenvalues depending on the distance. In particular, for the totally symmetric Wilson loop in ${\cal N}=4$, by taking into consideration the interaction among eigenvalues we improved the standing of the field theory against the holographic prediction. First, the corrected field theory result has the same sign as the holographic expression. Second, our corrected expression in the large $\kappa=k\sqrt{\lambda}/(4N)$ regime, is $-\log\kappa^2$ which is twice the holographic prediction up to an additive numerical constant.  It is important to compare more systematically our approach with different expansions including the large $\lambda$ expansions presented  \cite{Horikoshi:2016hds,Chen-Lin:2016kkk}. There has recently been a particularly enlighting clarification of the order of limits and corrections for the anti-symmetric Wilson loop in \cite{Gordon:2017dvy} \cite{Okuyama:2017feo} and it is expected that our work will contribute to a similar elucidation in the case of the symmetric representation.

There are a number of directions that would be interesting to explore. One natural direction pertains the gravitational counterpart of our computations. Namely, the construction of geometries corresponding to Wilson loops in large rectangular representations in ${\cal N}=2^*$. At first sight the task seems daunting as it involves constructing  a set of bubbling geometries such as those constructed in \cite{DHoker:2007mci} but on the background of the Pilch-Warner solution \cite{Pilch:2000ue}. The intricate structure of the solutions in  \cite{DHoker:2007mci} has been connected to the evaluation of Wilson loops in arbitrary representations in the beautiful analysis of  \cite{Okuda:2008px}. The relative  simplicity and universality of the Wilson loop expectation values we have obtained in this manuscript give hope that the construction of the fully back-reacted solution might be within reach.

On the gravity side, in the context of ${\cal N}=4$ holography, the higher dimensional representations have been explored beyond the leading term in a series of work involving the one-loop effective actions of D3 branes  \cite{faraggi:2011bb,Buchbinder:2014nia} and D5 branes \cite{Faraggi:2011ge}. The classical configuration discussed in \cite{Chen-Lin:2015xlh} constitutes a first step in the direction of being able to compare corrections on the holographic and field theory sides. A hopeful sign that the situation might be clearer in this case, despite the field theory being more complicated, is the recent positive result at the one-loop level for the fundamental representation in \cite{Chen-Lin:2017pay}.  Similarly, it will be very interesting to develop the status of higher rank representations Wilson loops in the context of ABJM theory . Indeed, the original work of \cite{Drukker:2008zx} introduced the holographic dual of the Wilson loops in higher rank representations of ABJM theory and used the field theory matrix model to compute the leading terms. More recently, the field theory computation has been carried beyond the leading term in \cite{Cookmeyer:2016dln} and the sub-leading structure on the gravity side was clarified in \cite{Muck:2016hda}. It seems only natural to elucidate the status of Wilson loops in rectangular representations of the ABJM theory. Another interesting direction involves considering other  deformations of ${\cal N}=4$ and ${\cal N}=2$  theories such as those presented in  \cite{Fucito:2015ofa} where some results for simple Wilson loops were also presented.

\section*{Acknowledgments}
We thank J. Aguilera-Darmia, X. Chen-Lin, D. Correa, F. Fucito, V. Giraldo-Rivera ,  J. F. Morales and G. Silva for various comments and discussions on closely related topics. This work is partially supported by the US Department of Energy under Grant No.\ DE-SC0007859 and Grant No. \ DE-SC0017808 -- {\it Topics in the AdS/CFT Correspondence: Precision tests with Wilson loops, quantum black holes and dualities}.

\appendix

\section{Genus one corrections}\label{App:genus}
In this appendix we include the material needed to support our results regarding sub-leading corrections to the expectation values of the Wilson loops discussed in the main text. 

 As an illustrative example we worked out explicitly the first correction to the Wilson loop (fundamental representation) in the Gaussian matrix model for arbitrary $\lambda$.

Using orthogonal polynomials, the Wilson loop in fundamental representation can be evaluated exactly \cite{Drukker:2000rr}
\begin{equation}
	\left<W_\square\right>\equiv\left<\frac{1}{N}\Tr\exp M\right>=\frac{1}{N}\sum_{j=0}^{N-1}L_j(-\lambda/4N)e^{\lambda/8N},
\end{equation}
where $L_j$ is the Laguerre polynomial. This Wilson loop admits a topological expansion of the form
\begin{equation}
	\left<W_\square\right>=\frac{2}{\sqrt{\lambda}}I_1(\sqrt{\lambda})+\frac{\lambda}{48N^2}I_2(\sqrt{\lambda})+\dots.
\end{equation}
In the following it will be shown how to reproduce this expansion using the loop function.

The loop function is defined as
\begin{equation}
	W_1(p)\equiv\left<\frac{1}{N}\Tr\frac{1}{p-M}\right>=\frac{1}{N}\sum_{k=0}^\infty\frac{\left<\Tr M^k\right>}{p^{k+1}},
\end{equation}
and admits a topological expansion
\begin{equation}
	W_1(p)=\sum_{g=0}^\infty\frac{1}{N^{2g}}W_1^{(g)}(p).
\end{equation}
Therefore in order to determine the Wilson loop we only need to expand the loop function at $p=\infty$ and replace $p^{k+1}$ with $k!$.

There is a systematic method to calculate $W_g(p)$ order by order \cite{Eynard:2005}. Define the $k$-point loop function as
\begin{equation}
	W_k(x_1,\dots,x_k)=N^{k-2}\left<\tr\frac{1}{	x_1-M}\dots\tr\frac{1}{x_k-M}\right>_c,
\end{equation}
where $c$ means connected part or cumulant, admitting a topological expansion
\begin{equation}
	W_k=\sum_{g=0}^\infty W_k^{(g)}.
\end{equation}
There is a recursion relation among the loop functions
\begin{equation}
\begin{split}
	2\sum_{m=0}^hW_1^{(h-m)}(x_1)W_k^{(m)}(x_1,\dots,x_k)+&\\
	+W_{k+1}^{(h-1)}(x_1,\dots,x_{h-1})+&\\
	+\sum_{m=0}^{h}\sum_{j=1}^{k-2}\sum_{I\in K_j}W_{j+1}^{(m)}(x_1,x_I)W_{k-j}^{(h-m)}(x_1,x_{K-I})+&\\
	+\sum_{j=2}^k\frac{\partial}{\partial x_j}\frac{W_{k-1}^{(h)}(x_2,\dots,x_j,\dots,x_k)-W_{k-1}^{(h)}(x_2,\dots,x_1,\dots,x_k)}{x_j-x_1}=&\\
	=V^\prime(x_1)W_k^{(h)}(x_1,\dots,x_k)-U_k^{(h)}(x_1;x_2,\dots,x_k).
\end{split}
\end{equation}
where $V$ is the potential of matrix model (for the Gaussian matrix model it is just a quadratic function), $K_j\equiv\{I\subset K=\{2,\dots,k\}|I|=j\},x_I\equiv\prod_{i\in I}x_i$ and
\begin{equation}
	U_k(x_1;x_2,\dots,x_k)\equiv N^{k-2}\left<\tr\frac{V_1^\prime(x_1)-V_1^\prime(M)}{x_1-M}\tr\frac{1}{x_2-M}\dots\tr\frac{1}{x_k-M}\right>_c	.
\end{equation}
When $k=1,h=1$, the recursion relation reduces to
\begin{equation}
	\sqrt{\sigma(x_1)}W_1^{(1)}(x_1)=\frac{W_2^{(0)}(x_1,x_1)+U_1^{(1)}(x_1)}{M(x_1)}
\end{equation}
where the functions $M(x)$ and $\sigma(x)$ come from the one loop function
\begin{equation}
	W_1^{(0)}(x)=\frac{1}{2}(V^\prime(x)-M(x)\sqrt{\sigma(x)}).
\end{equation}
The two-point function can be derived from the recursion relation ($k=1,h=0$)
\begin{equation}
	M(x_1)\sqrt{\sigma(x_1)}W_2^{(0)}(x_1,x_2)=\frac{\partial}{\partial x_2}\left(\frac{W_1^{(0)}(x_2)-W_1^{(0)}(x_1)}{x_2-x_1}\right)+U_2^{(0)}(x_1;x_2).
\end{equation}

In the Gaussian matrix model, the one-loop function can be derived either from the recursion relation or directly from the Wigner distribution
\begin{equation}
	W_1^{(0)}(p)=\frac{2}{\lambda}(p-\sqrt{p^2-\lambda}),
\end{equation}
and $\sigma(x)=p^2-\lambda, M(x)=\frac{4}{\lambda}$. The function $U$ can be evaluated by definition
\begin{equation}
	U_2(x_1)=\frac{1}{N}\left<\tr\frac{V^\prime(x)-V^\prime(M)}{x_1-M}\right>=1,
\end{equation} 
\begin{equation}
	U_2(x_1;x_2)=\left<\tr\frac{V^\prime(x)-V^\prime(M)}{x_1-M}\tr\frac{1}{x_2-M}\right>=NW_1^{(0)}(x_2),
\end{equation}
hence the two-point function is
\begin{equation}
	W_2^{(0)}(p,p)=\frac{4\lambda}{16(p-\sqrt{\lambda})^2(p+\sqrt{\lambda})^2},
\end{equation}
and the one-loop function is
\begin{equation}
	W_1^{(1)}(p)=\frac{1}{\sqrt{p^2-\lambda}}\frac{(\lambda/4)^2}{(p^2-\lambda)^2}.
\end{equation}

Expand the loop functions w.r.t. $p$, we have
\begin{equation}
	W_1^{(0)}(p)=\frac{2p}{\lambda}\left[1-\sum_{k=0}^\infty\binom{1/2}{k}\left(-\frac{\lambda}{p^2}\right)^k\right]=-\frac{2}{\lambda}\sum_{k=0}^\infty\binom{1/2}{k}\frac{(-\lambda)^k}{p^{2k-1}},
\end{equation}
\begin{equation}
	W_1^{(1)}(p)=\frac{\lambda^2}{16p^5}\sum_{k=0}^\infty\binom{-5/2}{k}\left(-\frac{\lambda}{p^2}\right)^k=\frac{\lambda^2}{16}\sum_{k=0}^\infty\binom{-5/2}{k}\frac{(-\lambda)^k}{p^{2k+5}}.
\end{equation}
The Wilson loop at leading order is obtained by the substitution $p^{k+1}\mapsto k!$
\begin{equation}
	-\frac{2}{\lambda}\sum_{k=0}^\infty\binom{1/2}{k}\frac{(-\lambda)^k}{(2k-2)!}=\frac{2}{\sqrt{\lambda}}I_1(\sqrt{\lambda}),
\end{equation}
and similarly the sub-leading order is given by
\begin{equation}
	\frac{\lambda^2}{16}\sum_{k=0}^\infty\binom{-5/2}{k}\frac{(-\lambda)^k}{(2k+4)!}=\frac{\lambda}{48}I_2(\sqrt{\lambda}),
\end{equation}
which matches precisely with the exact result.

\section{Further remarks on the antisymmetric representation}\label{App:Ak}

In this appendix we review the computations of the expectation value of the $k$-antisymmetric Wilson loop in ${\cal N}=4$ SYM, a similar treatment can be extended to ${\cal N}=2^*$. We would like to scrutinize the computation of \cite{Hartnoll:2006is} and the  set of approximations made in that work.




Recall that the anti-symmetric Wilson loop is given by
\begin{equation}\label{AntiSymWL}
	\left<W_{A_k}\right>=\frac{1}{\mathcal{Z}}\int_{0}^{2\pi i} du\int_{\mathbb{R}^N}\prod_{i=1}^N dm_i\exp\left(S[M]+n u + \sum_{i=1}^N\log\left(1-e^{m_i-u}\right)\right),
\end{equation}
where $S[M]$ is the action for $\mathcal{N}=4$ SYM (the functional form of $S[M]$ is not crucial, we just take the $\mathcal{N}=4$ theory for simplicity) which in terms of eigenvalues reads
\begin{equation}
	-\frac{2N}{\lambda}\sum_{i=1}^N m_i^2+\sum_{i<j}\log (m_i-m_j)^2\sim O(N^2).
\end{equation}
Notice that the integrand is periodic ($u\to u+2\pi i$) so it is better to think of $u$ on a cylinder.

The authors of \cite{Hartnoll:2006is} claim that there are poles at $u=m_i$ (or a cut at $[-\sqrt{\lambda},\sqrt{\lambda}]$) and we can choose a contour homotopic to $[-\infty,-\infty+2\pi i]$ for $u$ such that it passes through the saddle-point. However, it can be verified that the real part of the exponent in the integrand is minimized at the saddle-point and hence the integral is not dominated by its value at the saddle-point. Therefore, the calculation in \cite{Hartnoll:2006is} needs to be re-interpreted.

On the other hand, the solution for our saddle-point equation (\ref{SaddlePointEquationU1}) and (\ref{SaddlePointEquationU2}) is not unique. In fact, apart from the saddle-point we found in (\ref{AntiSymmetricSaddlePoint}) (denoted by $z_0$), $u$ can be located between any two consecutive eigenvalues $m_i$ or at $z_0+i\pi$, and the reason to choose $z_0$ requires a justification.

First of all, notice that after taking the exponential, there is actually not a pole or a cut at $u=m_i$, which implies that we can freely deform the contour for $u$, as long as it belongs to the same homotopy class. Hence we can simply choose the contour $[z_0,z_0+2\pi i]$ and safely ignore the other saddle-point configurations.

\begin{figure}[t]
  \centering
  \subfigure[$\Re w(u)$]{\label{ContourPlotRe}
    \label{fig:subfig:a} 
    \includegraphics[width=2in]{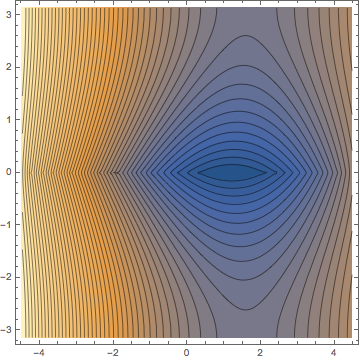}}
  \hspace{1in}
  \subfigure[$\Im w(u)$]{\label{ContourPlotIm}
    \label{fig:subfig:b} 
    \includegraphics[width=2in]{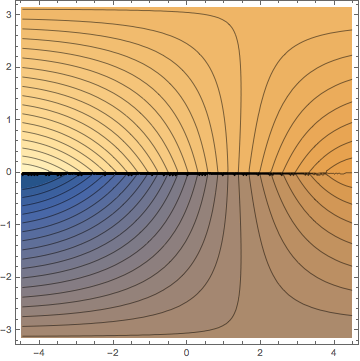}}
  \caption{Contour lines of $w(u)$}
 \label{Fig:contour}  
\end{figure}

Now we need to explain what happens if we choose other contours, since the integral should be independent of the contour. In order to employ saddle-point approximation, the contour cannot be arbitrary. Actually when we apply the so called saddle-point approximation to such a integral
\begin{equation}
	I(N)=\int dx g(x)\exp\left[N(f(x))\right],
\end{equation}
either steepest descend formula ($\Im f(x)$ is fixed) or stationary phase formula ($\Re f(x)$ is fixed) is employed. Therefore the choice of contour should keep either the real part or the imaginary part of
\begin{equation}
	w(u)\equiv n u+\sum_{i=1}^N\log\left(1-e^{m_i-u}\right)
\end{equation}
constant. In general the contour line of $\Re w(u)$ and $\Im w(u)$ are as in Fig. \ref{Fig:contour}.

It is evident that the only steepest descent contour is $[z_0,z_0+2\pi i]$. If a stationary phase contour is chosen, then from Figure \ref{ContourPlotIm} we know there is no stationary phase point and we need to collect the contribution coming from the whole contour, where the rapid oscillation will cancel the larger amount of real part. Let us return to the unique steepest descent contour $[z_0,z_0+2\pi i]$, along which $\Re w(u)$ is maximized at $z_0+i \pi$. Therefore the integral (\ref{AntiSymWL}) is dominated by the saddle-point $z_0+i \pi$. In the large $N$ and strong-coupling limit $\lambda>>1$, the difference between the two saddle-point $z_0$ and $z_0+i\pi$ can be ignored (in other words, $\Re w(u)$ along the contour is nearly constant) and hence our previous calculation for anti-symmetric Wilson loop is valid.

\bibliographystyle{JHEP}
\bibliography{WLoops-bib}

\end{document}